\documentclass[twoside,reqno]{heron}
\usepackage{epsfig,cite,colordvi}
\usepackage{graphicx}
\usepackage{url}
\usepackage{amssymb,amsmath,amscd,epsf}
\usepackage{times}
\usepackage{makeidx}
\usepackage{physics}
\usepackage[version=3] {mhchem}

\graphicspath{ {c:/Figures/} }

\newcounter{mycite}
\newtoks\citetoks
\makeatletter
\DeclareRobustCommand\unscite[1]{%
  \@ifundefined{uns@cite#1}
    {\refstepcounter{mycite}\label{citelabel@#1}%
     \expandafter\xdef\csname uns@cite#1\endcsname{\arabic{mycite}}%
     \toks\z@=\expandafter{\the\citetoks}%
     \toks\tw@=\expandafter\expandafter\expandafter{%
       \csname uns@bibitem#1\endcsname}%
     \edef\@tempcite{\the\toks\z@\the\toks\tw@}%
     \global\citetoks=\expandafter{\@tempcite}%
    }{}[\@nameuse{uns@cite#1}]}
\newcommand{\mybibitem}[2]{%
  \@namedef{uns@bibitem#1}{\bibitem[\ref{citelabel@#1}]{#1}#2}}
\makeatother

\mybibitem{Rowe}{D. J. Rowe, {\it Rep. Prog. Phys.} 48, 1419-1480, (1985).}
\mybibitem{review}{P. Van Isacker and S. Pittel, {\it Physics Scripta, Vol. 91, Number 2}, (2016).}
\mybibitem{SymplecticI}{G. Rosensteel and D.J. Rowe, {\it Annals of Physics {\bf 123}, 36-60},(1979).}
\mybibitem{SymplecticII}{D. J. Rowe and G. Rosensteel, {\it Annals of Physics {\bf 126}, 198-233}, (1980).}
\mybibitem{SymplecticIII}{G. Rosensteel and D.J. Rowe, {\it Annals of Physics {\bf 126}, 343-370},(1980).}
\mybibitem{Park}{P. Park, J. Carvalho, M. Vassanji and D.J. Rowe, {\it Nuclear Physics {\bf A414}, 93-112}, (1983). }
\mybibitem{Sofia}{D. Bonatsos, S. Karampagia, R. B. Cakirli, R. F. Casten, K. Blaum and Amon Susam, {\it Phys. Rev. C 88, 054309,} (2013).}
\mybibitem{calculation}{D. Bonatsos \etal, {\it Phys. Rev. C} {\bf 95}, 064325 (2017). }

\mybibitem{lambda-mu}{ J. P. Draayer \etal {\it Computer Physics Communications 56, 279-290,} (1989).}
\mybibitem{Nilssonbook}{S.G. Nilsson and I. Ragnarsson, {\it Shapes and Shells in Nuclear Structure}, Cambridge University Press, (1995).}
\mybibitem{dSU(3)}{D. Bonatsos, C. Daskaloyannis, P. Kolokotronis and D. Lenis, {\it arXiv: hep-th/9411218v1}, (1994).}
\mybibitem{wf}{V. X. Genest, L. Vinet, A. Zhedanov, {\it arXiv: 1307.0692v2, math-ph}, (2013). }
\mybibitem{h.w.}{D. Bonatsos \etal {\it arXiv: 1712.04126v1, nucl-th,} (2017).}
\mybibitem{beta}{O.Castanos \etal {\it Z. Phys. A} {\bf 329}, 33-43, (1988).}
\mybibitem{Harvey}  {M. Harvey, in {\it Advances in Nuclear Physics}, Vol. 1, (Plenum, New York, 1968). }
\mybibitem{Cohen}{C. Cohen-Tannoudji, B. Diu, F. Laloe, {\it Quantum mechanics Volume I}, (WILEY-VCH).}
\mybibitem{1Nilsson}{S. G. Nilsson,  Dan Mat Fys Medd {\bf 29}, no 16 (1955).}
\mybibitem{Bahri} {C. Bahri, J. Escher, J. P. Draayer, Nuclear Physics A 592 (1995) 171-193.}
\mybibitem{Jutta} {J. Escher, C. Bahri, D. Troltenier, J. P. Draayer, Nuclear Physics A 633 (1998) 662-680.}
\mybibitem{64}{J.M. Yao, Z.X. Li, J. Xiang, H. Mei, {\it arXiv: 1010.4372v1}, (2010).}
\mybibitem{QQ}{J. P. Draayer, K.J. Weeks and K. T. Hecht, {\it Nuclear Physics A381, 1-12} (1982).}
\mybibitem{Spheroid}{Wikipedia: Spheroid.}
\mybibitem{Evolution}{R. G. Casten Lectures at Riken, January 2010.}
\mybibitem{Magic}{O. Sorlin and M.G. Porquet, {\it Nuclear Magic Numbers: new features far from stability,} arXiv:0805.2561, (2008). }
\mybibitem{Casten} {R.F. Casten, {\it Nuclear Structure from a Simple Perspective}, (Oxford University Press Inc., New York, 1990).}
\mybibitem{Nuclear-Physics} {W.N. Cottingham and D. A. Greenwood, {\it \textgreek{Εισαγωγή στην Πυρηνική Φυσική}}, (\textgreek{τυπωθήτω-Γιώργος Δάρδανος, Αθήνα 1995}).}
\mybibitem{Greiner}{ W. Greiner and J. A. Maruhn, {\it Nuclear Models} (Springer, Berlin, 1996).}
\mybibitem{Casten} {R.F. Casten, {\it Nuclear Structure from a Simple Perspective}, (Oxford University Press Inc., New York, 1990).}
\mybibitem{Bible} {J. P. Draayer \etal {\it Algebraic Approaches to Nuclear Structure}, (Harwood, 1993). }
\mybibitem{W}{ A. R. Edmonds, {\it Angular Momentum in Quantum Mechanics,} (Princeton University Press, Princeton, 1957).}
\mybibitem{Lipkin} {Harry J. Lipkin, {\it Lie groups for pedestrians}, (Dover Publications, 2002).}
\mybibitem{Nuclear-Physics} {W.N. Cottingham and D. A. Greenwood, {\it \textgreek{Εισαγωγή στην Πυρηνική Φυσική}}, (\textgreek{τυπωθήτω-Γιώργος Δάρδανος, Αθήνα 1995}).}

\mybibitem{book} {S. G. Nilsson, I. Ragnarsson, {\it Shapes and shells in nuclear structure,} (Cambridge University Press, 1995.)}
\mybibitem{Berkeley}{Robert Littlejohn, Lecture notes on Quantum Mechanics, University of Berkeley, (2010). }
\mybibitem{IBM}{F. Iachello, A. Arima, {\it The Interacting boson model}, (Cambridge University Press, 1987).}
\mybibitem{applet}{ http://mathinsight.org/determinant\_linear\_transformation.}
\mybibitem{Wigner-Eckart}{ J. P. Draayer and Yoshimi Akiyama, {\it J. of Math. Phys.,  Vol 14,  No 12}, (1973).}
\mybibitem{rigid-rotator}{ O. Casta$\tilde n$os, J. P. Draayer and Y. Leschber, {\it Z. Phys. A - Atomic Nuclei 329, 33-43,} (1988).}
\mybibitem{Naqvi} {H. A. Naqvi and J. P. Draayer, {\it Nuclear Physics A516, 351-364,} (1990).}
\mybibitem{Gelfand}{ Jes$\acute{u}$s A. De Loera \etal {\it Discrete Comput. Geom. {\bf 32} 459-470,} (2004).}
\mybibitem{lambda-mu}{ J. P. Draayer \etal {\it Comp. Phys. Com. {\bf 56}, 279-290,} (1989).}
\mybibitem{Wood}{J.L. Wood \etal {\it Phys.  Rep. {\bf 215}, 101-201} (1992).}
\mybibitem{coexistence}{ K. Heyde and J. L. Wood, {\it Rev. Mod. Phys. {\bf 83}}, 1467-1521 (2011).}
\mybibitem{CG-coefficients} {C. Bahri, D. J. Rowe, J. P. Draayer, {\it Computer Physics Communications 159, 121-143}, (2004).}
\mybibitem{Elliotta}{ J. P. Elliott, {\it Proc. Roy. Soc. Ser. A {\bf 245}, 128-145}, (1958).}
\mybibitem{Elliottb}{ J. P. Elliott, {\it Proc. Roy. Soc. Ser. A {\bf 245}, 562-581}, (1958).}
\mybibitem{Elliottc}{ J. P. Elliott and M. Harvey, {\it Proc. Roy. Soc. Ser. A {\bf 272}, 557-577}, (1963).}
\mybibitem{Q} {J.P. Draayer and K. J. Weeks, {\it Annals of Physics, 156, 41-67}, (1984).}
\mybibitem{TE2}{ O. Casta$\tilde n$os, J. P. Draayer and Y. Leschber, {\it Annals of Physics 180, 290-329}, (1987).}
\mybibitem{ENSDF}{ Brookhaven National Laboratory ENSDF database, http://www.nndc.bnl.gov/ensdf/}
\mybibitem{W.U.}{ S. Raman, C.W. Nestor, Jr., and P. Tikkanen, {\it  At. Data Nucl. Data Tables 78, 1},
(2001).}
\mybibitem{Sofia}{D. Bonatsos, S. Karampagia, R. B. Cakirli, R. F. Casten, K. Blaum and Amon Susam, {\it Phys. Rev. C 88, 054309,} (2013).}
\mybibitem{pn pairs} {Dennis Bonatsos, I.E. Assimakis and A. Martinou, {\it Bulg. J. Phys. 42, 439-449}, (2015).}
\mybibitem{Rila} {Andriana Martinou, I.E. Assimakis, N. Minkov, D. Bonatsos, {\it, Nuclear Theory '35, Proceedings of the 35th  International Workshop on Nuclear Theory (Rila 2016),  224-235}, (Heron Press, Sofia, 2016).}

\mybibitem{RilaSmaragda}{S.  Sarantopoulou, A.Martinou, I. E. Assimakis, N. Minkov, and D. Bonatsos, {\it , Nuclear Theory '35, Proceedings of the 35th  International Workshop on Nuclear Theory (Rila 2016), 236-243,} (Heron Press, Sofia, 2016).}
 
\mybibitem{po} {D. Bonatsos \etal {\it Phys. Rev. C} {\bf 95}, 064326, (2017).}

\mybibitem{SofiaAndriana}{A. Martinou, D. Bonatsos, I. E. Assimakis, N. Minkov, S. Sarantopoulou, R. B. Cakirli, R. F. Casten, and K. Blaum,{\it  Bulg. J. Phys.  (2017) in press.  Proceedings of the Workshop on ``Shapes and Dynamics of Atomic Nuclei: Contemporary Aspects’’ (SDANCA17, Sofia 2017), ed. N. Minkov}}

\mybibitem{SofiaBonatsos}{D. Bonatsos, I. E. Assimakis, N. Minkov, A. Martinou, S. K. Peroulis, S. Sarantopoulou, R. B. Cakirli, R. F. Casten, and K. Blaum, {\it Bulg. J. Phys.  (2017) in press.  Proceedings of the Workshop on ``Shapes and Dynamics of Atomic Nuclei: Contemporary Aspects’’ (SDANCA17, Sofia 2017), ed. N. Minkov}}

\mybibitem{SofiaSmaragda}{S. Sarantopoulou, D. Bonatsos, I. E. Assimakis, N. Minkov, A. Martinou, R. B. Cakirli, R. F. Casten, and K. Blaum, {\it Bulg. J. Phys.  (2017) in press.  Proceedings of the Workshop on ``Shapes and Dynamics of Atomic Nuclei: Contemporary Aspects’’ (SDANCA17, Sofia 2017), ed. N. Minkov}}

\mybibitem{SofiaGiannis}{I. E. Assimakis, D. Bonatsos, N. Minkov, A. Martinou,  R. B. Cakirli, R. F. Casten, and K. Blaum, {\it Bulg. J. Phys.  (2017) in press.  Proceedings of the Workshop on ``Shapes and Dynamics of Atomic Nuclei: Contemporary Aspects’’ (SDANCA17, Sofia 2017), ed. N. Minkov.}}

\mybibitem{Kr}{ F. Flavigny et al., {\it PRL 118, 242501}, (2017).}
\mybibitem{irreps} {P. Van Isacker, D. D. Warner and D. S. Brenner, {\it Phys. Rev. Let. {\bf 74}, 23}, (1995).}
\mybibitem{182Hg} {N. Bree et al. {\it Phys. Rev. Lett. 112, 162701}, (April 2014).}
\mybibitem{moments}{T.J. Mertzimekis, K. Stamou, A. Psaltis, Nucl. Instrum. Meth. Phys. Res. A 807, 56 (2016).}
\mybibitem{Ge}{Sugawara, M., Toh, Y., Czosnyka, T. et al. Eur Phys J A (2003) 16: 409.}
\mybibitem{Clement} {E. Cl\'ement et al., {\it Phys. Rev. C 75, 054313,} (2017).}
\mybibitem{70Se}{ A. M. Hurst et al., {\it Phys. Rev. Lett. 98, 072501}, (2007).}
\mybibitem{Asherova}{R. M. Asherova, Yu. F. Smirnov, V. N. Tolstoy, {\it Nuclear Physics, A355, 25-44}, (1981).}
\mybibitem{HC}{K. Heyde \etal \textit{Phys. Lett. {\bf 155B}, 303-308}, (1985)}
\mybibitem{Sorlin}{O. Sorlin and M.-G. Porquet, {\it Prog. Part. Nucl. Phys.} {\bf 61}, 602-673 (2008).}

\begin{document}

\title{Highest weight SU(3) irreducible representations for nuclei with shape coexistence}

\runningheads{Highest weight SU(3) irreps for nuclei with shape coexistence}{A. Martinou, D. Bonatsos, N. Minkov, I. E. Assimakis, S. Sarantopoulou, {\it et al.}}

\begin{start}

\author{Andriana Martinou}{1}, \coauthor{Dennis Bonatsos}{1}, \coauthor{N. Minkov}{2}, \coauthor{I.E. Assimakis}{1}, \coauthor{S. Sarantopoulou}{1}, \coauthor{S. Peroulis}{3}

\address{Institute of Nuclear and Particle Physics, National Centre for Scientific Research  ``Demokritos'', GR-15310 Aghia Paraskevi, Attiki, Greece}{1}

\address{Institute of Nuclear Research and Nuclear Energy, Bulgarian Academy of Sciences, 72 Tzarigrad Road, 1784 Sofia, Bulgaria}{2}

\address{University of Athens, Faculty of Physics, Zografou Campus, GR-15784 Athens, Greece}{3}

\begin{Abstract}
The SU(3) irreducible representations (irreps) are characterised by the $(\lambda, \mu)$ Elliott quantum numbers, which are necessary for the extraction of the nuclear deformation, the energy spectrum and the transition probabilities. These irreps can be calculated through a code which requires high computational power. In the following text a hand-writing method is presented for the calculation of the highest weight (h.w.) irreps, using two different sets of magic numbers, namely proxy-SU(3) and three-dimensional isotropic harmonic oscillator.
\end{Abstract}
\end{start}

\section{Introduction}

The exact interaction among two individual nucleons is not yet known. Therefore effective interactions are used in the variety of nuclear models, which all aim to solve the nuclear many body problem. These interactions can be sorted into three categories. The first one is the interaction of a nucleon with the mean field created by the rest of the nucleons in the system. This mean field in the SU(3) models is represented by a three dimensional (3D) isotropic harmonic oscillator (HO) potential. The second category is the quadrupole-quadrupole ($Q\cdot Q$) interaction, which is used for deformed nuclei and it is the long range part of the nucleon-nucleon potential \unscite{Harvey}. The third category is the pairing force. The Elliott SU(3) model \unscite{Elliotta}, \unscite{Elliottb}, \unscite{Elliottc} is using the 3D isotropic HO potential as the leading interaction and the quadrupole-quadrupole ($Q\cdot Q$) interaction as a secondary term, which creates deformation. Hereafter the capital
letter $Q$ will be used for the whole nucleus, while $q$ will
denote the single particle operator.

The SU(3) algebra has two quantum numbers, namely $(\lambda, \mu)$. These numbers are sufficient for defining the eigenvalue of the SU(3) second order Casimir operator (par. 7.2.3. of \unscite{Bible})
\begin{equation}\label{C}
<C_{SU(3)}^{(2)}>=\lambda ^2+\mu^2+\lambda \mu+3(\lambda+\mu).
\end{equation}
This operator is related to the Bohr-Mottelson deformation variable $\gamma$ through \unscite{beta}
\begin{eqnarray}
\gamma =tan^{-1}({\sqrt{3}(\mu+1)\over 2\lambda +\mu+3}).\label{g}
\end{eqnarray}
Furthermore the Casimir operator is connected with the $Q\cdot Q$ interaction through (par. 7.1.5. of \unscite{Bible})
\begin{equation}\label{C}
Q\cdot Q=4C_{SU(3)}^{(2)}-3L^2,
\end{equation}
where $L$ is the angular momentum. A basic SU(3) Hamiltonian is
\begin{equation}\label{H}
H=H_0-{1\over 2}\chi Q\cdot Q,
\end{equation}
with $\chi$ being the strength of the $Q\cdot Q$ interaction and $H_0=\sum_{i=1}^A({p_i^2\over 2m}+{1\over 2}m\omega ^2r_i^2)$. Therefore the $(\lambda, \mu)$ quantum numbers are crucial for defining the deformation and the energy of the nucleus.

\section{Hand-writing method for the highest weight irreps of SU(3)}

The procedure for the calculation of SU(3) irreps involves Gel'fand-Tsetlin triangles and it is described very well in ref. \unscite{lambda-mu}. This reference is accompanied by a code, which reproduces all possible irreps. But if one wants simply the h.w. irrep for a nucleus without running the code, it is safe to use the following method.

Let $n=n_x+n_y+n_z$ be the total number of quanta for the isotropic 3D-HO problem. The eigenfunctions of the 3D isotropic HO problem, \ie $H=\sum _{i=z,x,y}({p_i^2\over 2m}+{1\over 2}m\omega ^2 x_i^2)$ in the Cartesian coordinate system are $\ket{n_z,n_x,n_y}$. For $n=3$, 10 orbitals of this type emerge (thus the corresponding algebra is U(10)): 
\begin{eqnarray*}
\ket{1}=\ket{3,0,0}, \ket{2}=\ket{2,1,0}, \ket{3}=\ket{2,0,1}, \ket{4}=\ket{1,2,0}, \ket{5}=\ket{1,1,1},\\ \ket{6}=\ket{1,0,2}, \ket{7}=\ket{0,3,0}, \ket{8}=\ket{0,2,1}, \ket{9}=\ket{0,1,2}, \ket{10}=\ket{0,0,3}.
\end{eqnarray*}
The order of the orbitals is very important for getting the right results. First comes the orbital with $n_z=n$, followed by the orbitals $\ket{2}, \ket{3}$, which  have $n_z=n-1$. Among them, $\ket{2}$ gains priority against $\ket{3}$, because it has more quanta in the $x$-axis. Similar rules apply to the orbitals $\ket{4}, \ket{5}, \ket{6}$ \etc. 
\subsection{The Gel'fand-Tsetlin triangle}

\begin{figure}
\begin{center}
\caption{A typical Gel'fand Tsetlin triangle. Along all the parallels of the red and green lines the numbers must not increase. For $k$ particles the h.w. irrep results if all the numbers along the first $k\over 2$ red parallels are 2 and along the rest red parallels are 0.}\label{Triangle}
\includegraphics[scale=0.2]{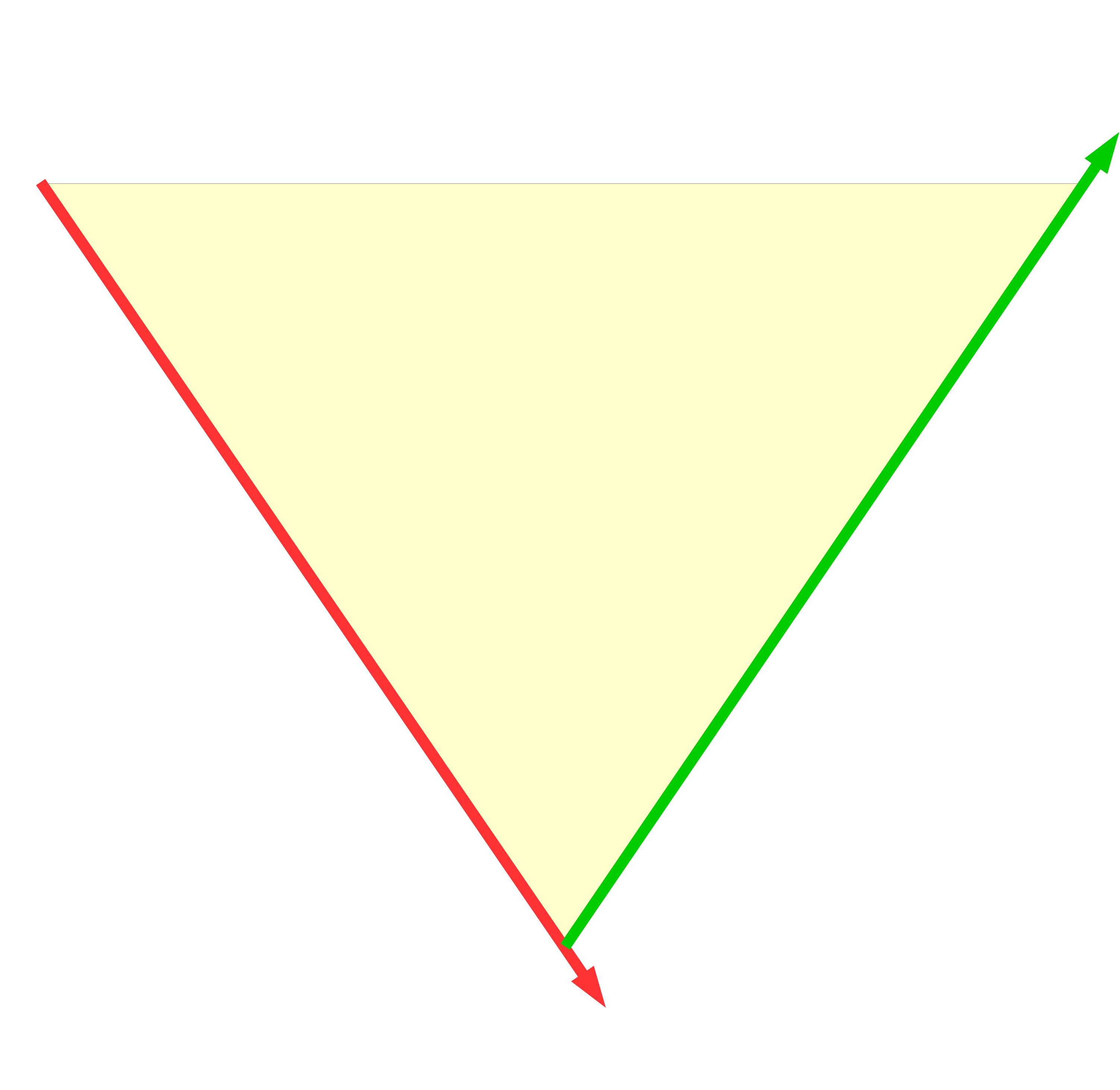}
\end{center}
\end{figure}

Each orbital can be occupied by two protons (neutrons). The question is how $k$ particles are distributed into $N$ orbitals? There is a variety of combinations, each of them leading to an irrep characterised by a set of $(\lambda, \mu)$ quantum numbers of SU(3). In many cases multiple different distributions result to the same irrep. Mathematically the way to explore all these distributions is the Gel'fand-Tsetlin pattern \unscite{Gelfand}. The pattern is simply a set of numbers, arranged in a triangular shape, following a basic rule: {\it the numbers must not increase along all the red and green parallels of Fig.(\ref{Triangle})}. Each combination of numbers that does satisfy this simple rule corresponds to a particle distribution.

For the U(10) algebra the Gel'fand-Tsetlin (GT) triangle is an equilateral triangle, filled with 10 numbers on each side ( for \eg Fig. (\ref{GT1})). These numbers can be 0, 1 or 2 since each orbital can be occupied with up to 2 protons (neutrons). Let start counting the horizontal lines $i$ (with $i=1,2,...,N$ for a U(N) algebra) of the triangle from {\it bottom to top}, so as the $i^{th}$ line to be filled with $i$ numbers. Let also $S_i$ to be the sum of numbers in each line, which equals to the number of nucleons that have been placed up to the $i^{th}$ orbital. For instance $S_4$ shows how many nucleons have been placed in orbitals $\ket{1},...,\ket{4}$. For $k$ valence protons (neutrons) the first $k\over 2$ numbers of the $N^{th}$ horizontal line are always 2, while the rest numbers of the same line are always 0 (\eg Fig. (\ref{GT1})-Fig. (\ref{GT6})). The previous horizontal lines $i=1,2,...,N-1$ are filled with numbers so as along all the red and green parallels of Fig. (\ref{Triangle}) the numbers do not increase. 

For the h.w. irrep the numbers along the first $k\over 2$ red parallels of Fig. (\ref{Triangle}) are all equal to 2 and along the rest red parallels are equal to 0 (\eg Fig.(\ref{GT1})). The weight vector is $\vec w=(w_1,w_2,...,w_{10})$. By definition $w_1=S_1, w_2=S_2-S_1, w_3=S_3-S_2, \etc$ Each of these $w_i$ is the occupation number of the $i^{th}$ orbital. Highest weight irrep has the maximum possible occupation number ($w_i=2$) for the first $k\over 2$ orbitals. 

\subsection{Proxy highest weight irreps}

Over the years there has been a doubt on which irrep reflects to the ground state band. The two candidates are the h.w. irrep and the one with the largest eigenvalue of $C^{(2)}_{SU(3)}$, oftenly called leading irrep. From one side the eigenvalue of the $C^{(2)}_{SU(3)}$ operator in eq.(\ref{C}) is a measure of $Q\cdot Q$ interaction, which minimises the energy. Therefore the argument is that the irrep with the greatest eigenvalue of $C^{(2)}_{SU(3)}$ gives the lowest energy. On the other side research on binding energies revealed, that the short range character of the nuclear force favors the irrep with the most symmetric spatial wave function \unscite{irreps}. This irrep proves to be the h.w. \unscite{h.w.} . Furthermore this irrep predicts right the prolate-oblate shape phase transition point \unscite{po}.

So the question {\it what is the meaning of the highest weight irrep} arises. The h.w. irrep is a specific filling order of the orbitals with nucleons. The orbitals are characterised by the $n_\perp=n_x+n_y$ quantum number. Each set with specific $n_\perp$ is created by $n_\perp+1$ orbitals. For instance three orbitals ($\ket{4}, \ket{5}, \ket{6}$) have $n_\perp=2$. 

The filling order can be better understood in two steps:\\
First are sorted the orbitals with increasing value of $n_\perp=0,1,2,...,n$. This favors the orbitals with maximum eigenvalue of the $m=0$ component of the quadrupole moment spherical tensor \ie  $q_0=2n_z-n_\perp$. In such a case the length of the $z$ axis is preferred to be maximum, when compared to that of the other two axes. So it favors the prolate cylindrical shape. \\
Secondly inside each block with specific $n_\perp$ priority is given to one axis, let it be named $x$ (the axes $x$ and $y$ in cylindrical symmetry are indistinguishable), so as the length of the $x$ axis to be maximum, when compared to that of the $y$ axis. This ordering favors triaxiality. In general the h.w. irrep favors deformation. 

The proxy-SU(3) symmetry has been recently proposed in refs. \unscite{po}, \unscite{calculation}. An approximation is made, in order to restore the SU(3) symmetry, destroyed by the spin-orbit ($l\cdot s$) interaction. Because of this approximation, the intruder orbitals invading a nuclear shell from above are replaced by orbitals bearing similar quantum numbers, which however have one less quantum in the $z$-axis. The resulting proxy-SU(3) shells are 28-48, 50-80, 82-124,...

\begin{figure}
\begin{center}
\caption{GT triangle for 6 valence particles in U(10), for the h.w. irrep. Along the first ${6\over 2}=3$ red parallels of Fig. (\ref{Triangle}) all the numbers are 2, while in the rest red parallels the numbers are 0. This placement results to the h.w. irrep.}\label{GT1}
\bigskip
\includegraphics[scale=0.7]{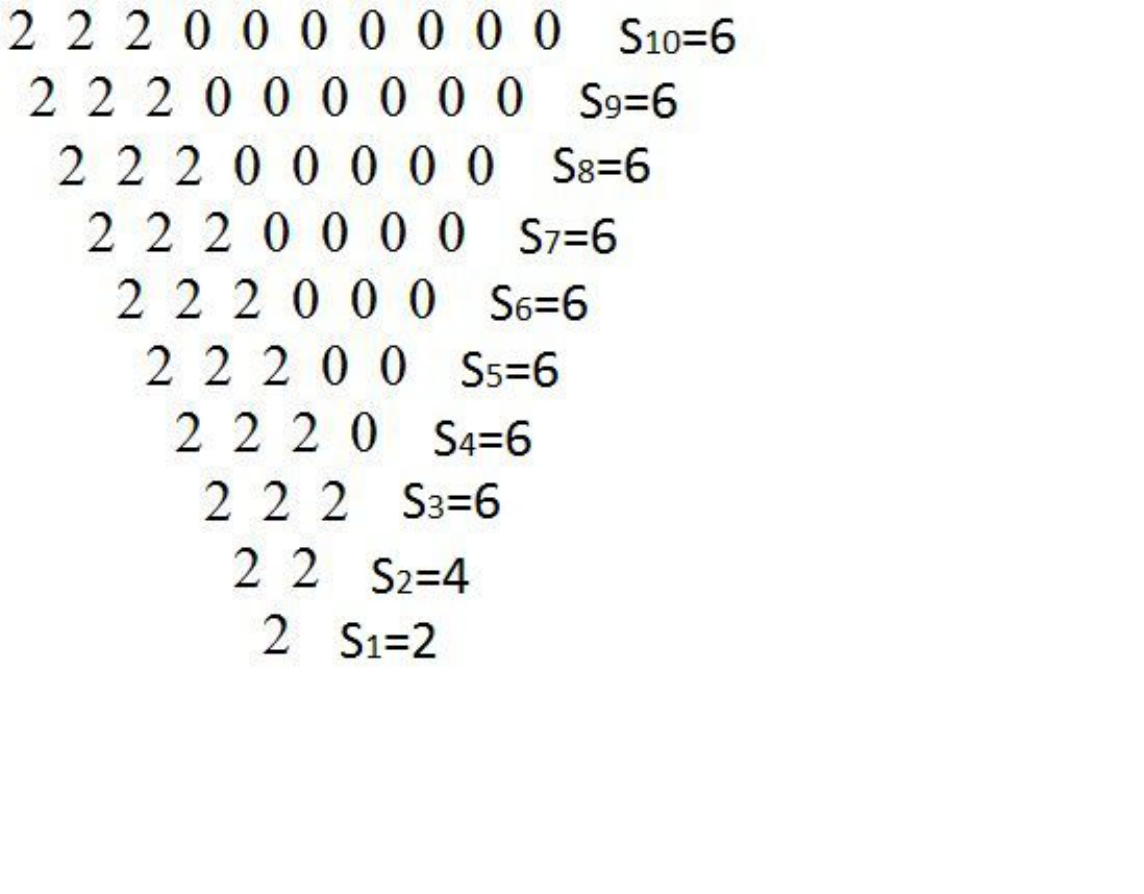}
\end{center}
\end{figure}

As an example, the calculation for $\ce{^{72}_{34}Se_{38}}$ will be presented. The 34 protons lie in the shell 28-48, thus this nucleus has 6 valence protons. The corresponding GT triangle is on Fig. (\ref{GT1}). Thus $\vec w=(2,2,2,0,0,0,0,0,0,0)$. This means that the orbitals $\ket{1}, \ket{2}, \ket{3}$ are occupied with 2 protons each, while the rest are empty. The next step is to sum the number of quanta in each of the three Cartesian axes: $\sum n_z=2\cdot 3+2\cdot 2+2\cdot 2=14$, $\sum n_x=2\cdot 0+2\cdot 1+2\cdot 0=2$ and $\sum n_y=2\cdot 0+2\cdot 0+2\cdot 1=2$. The results $(14,2,2)$ have to be arranged in decreasing order, in order to represent the quantum numbers $[f_1,f_2,f_3]$ of the U(3) symmetry, \ie $[f_1f_2f_3]=[14, 2, 2]$. The last step is to use the relations
\begin{equation}
\lambda=f_1-f_2, \mu=f_2-f_3.
\end{equation} 
Finally the proton irrep is $(\lambda_{1p}, \mu_{1p})=(12,0)$.

\begin{figure}
\begin{center}
\caption{GT triangle for 10 valence particles in U(10), for the h.w. irrep.}\label{GT2}
\bigskip
\includegraphics[scale=0.6]{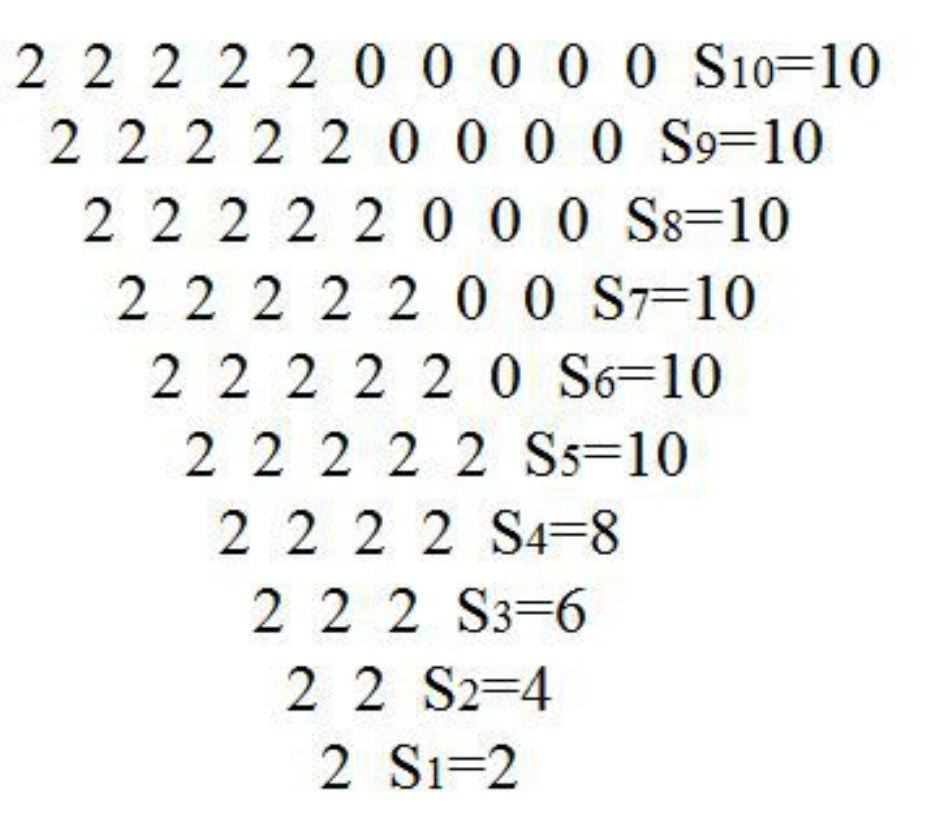}
\end{center}
\end{figure}
The 38 neutrons lie again in the 28-48 shell, which means that this nucleus has 10 valence neutrons. The corresponding GT triangle is Fig. (\ref{GT2}). Thus the weight vector is $\vec w=(2,2,2,2,2,0,0,0,0,0)$. So the valence neutrons occupy the first five orbitals: $\ket{1}, \ket{2},...,\ket{5}$. The summation of quanta gives $\sum n_z=2\cdot 3+2\cdot 2+2\cdot 2+2\cdot 1+2\cdot 1=18$, $\sum n_x=2\cdot 0+2\cdot 1+2\cdot 0+2\cdot 2+2\cdot 1=8$ and $\sum n_y=2\cdot 0+2\cdot 0+2\cdot 1+2\cdot 0+2\cdot 1=4$. Therefore $[f_1,f_2,f_3]=[18,8,4]$. The neutron irrep is $(\lambda_{1N}, \mu_{1N})=(10,4)$.

Then the total proxy-SU(3) h.w. irrep is
\begin{equation}
(\lambda_1,\mu_1)=(\lambda_{1p}, \mu_{1p})+(\lambda_{1N}, \mu_{1N}),
\end{equation}
 which for $\ce{^{72}_{34}Se_{38}}$ is $(22,4)$. Using eq. (\ref{g}) for this irrep gives $\gamma=9.64^\circ$. Since this value is much lower than $30^\circ$, the nucleus in such a state is prolate.

\section{Comparison of the h.w. with the leading irrep}

In the previous example in $\ce{^{72}_{34}Se_{38}}$ it became clear that the 10 valence neutrons occupy orbitals $\ket{1},\ket{2}, \ket{3}, \ket{4},\ket{5}$. This irrep is simultaneously h.w. and leading, because it happened to have the greatest eigenvalue of the $C_{SU(3)}^{(2)}$.

\begin{figure}
\begin{center}
\caption{GT triangle for 12 valence particles in U(10), for the leading irrep.}\label{GT6}
\bigskip
\includegraphics[scale=0.5]{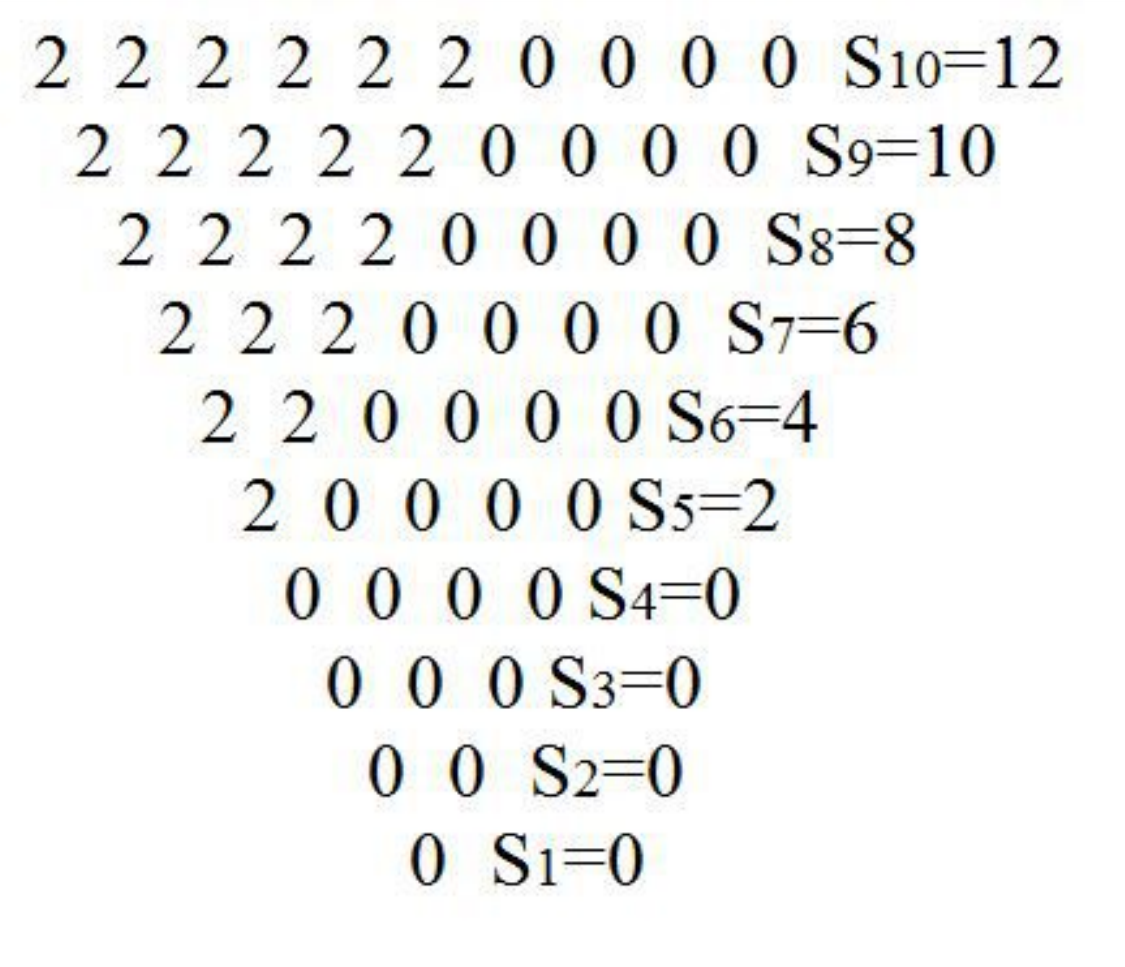}
\end{center}
\end{figure}

It is interesting to see what happens if 2 more neutrons are added: $\ce{^{74}_{34}Se_{40}}$. In this case the h.w. irrep is being derived if the 12 valence neutrons occupy the first 6 orbitals $\ket{1},...,\ket{6}$. The outcome is $(\lambda, \mu)=(12,0)$. The eigenvalue of the second order Casimir operator, as defined in eq. (\ref{C}) is:
\begin{equation*}
C_{SU(3)}^{(2)}=180, \mbox{ for h.w. irrep.}
\end{equation*}
But for 12 particles in U(10) there is another irrep (leading), which has a greater eigenvalue for the above operator. This irrep has $(\lambda, \mu)=(4,10)$ and
\begin{equation*}
C_{SU(3)}^{(2)}=198, \mbox{ for leading irrep.}
\end{equation*}
The weight vector of the corresponding GT triangle (Fig. (\ref{GT6})) is\\ $\vec w=(0,0,0,0,2,2,2,2,2,2)$, which means that the 12 particles are now distributed in orbitals $\ket{5},\ket{6}, \ket{7}, \ket{8}, \ket{9}, \ket{10}$. Following the procedure for the extraction of the irrep one gets that $\sum n_z=4,\sum n_x=14, \sum n_y=18$. These summations are an approximate measure for the length of each axis, since in the HO mean field potential using the Virial Theorem it is derived that $<x^2>={\hbar \over m\omega}(\sum n_x+{1\over 2})$. Thus the nucleus in this irrep has two axes ($x,y$) elongated, when compared to the $z$ axis. Obviously this state corresponds to an oblate shape. As a result the leading irrep predicts a prolate-oblate shape transition in the mid shell. But it is well known through experimental data that the prolate-oblate shape transition occurs after mid shell regions.

It is convenient now to compare the particle distributions of the leading irrep:
\begin{eqnarray*}
\ket{1},\ket{2}, \ket{3}, \ket{4},\ket{5}, \mbox{ 10 valence particles in U(10) for both h.w. and leading},\\
\ket{5},\ket{6}, \ket{7}, \ket{8}, \ket{9}, \ket{10}, \mbox{ 12 valence particles in U(10) for leading irrep}.
\end{eqnarray*}
It seems that, by following the leading irrep, the addition of the 2 more neutrons in orbital $\ket{6}$, causes a huge displacement of 8 particles from the orbitals $\ket{1},\ket{2}, \ket{3}, \ket{4}$ to the orbitals $ \ket{7}, \ket{8}, \ket{9}, $ $\ket{10}$. So the leading irrep is accompanied by tremendous particle rearrangements in the orbitals, every time 2 more particles are added in the shell. 

On the contrary the h.w. irrep is not causing any particle displacements. Since the 10 particles have already occupied the orbitals $\ket{1},...,\ket{5}$, the additional pair of 2 neutrons is being placed to one of the orbitals $\ket{6},...,\ket{10}$, which is furthermore compatible with the GT triangle. Among the possibilities which respect the GT pattern and cause no particle displacements of the 10 previous neutrons, the chosen one is that of {\it the greatest value of the $C_{SU(3)}^{(2)}$}. 

To conclude, although the leading irrep would result to a lower energy it seems that it is blocked by the previously placed particles. Indeed the prolate-oblate shape transition after the mid shell indicates, that the particles follow a smooth distribution in the orbitals, without serious rearrangements.

\section{Shape coexistence}

Shape coexistence appears when a nucleus has two low lying $0^+$ energy bands close to each other, each of them exhibiting a different shape. The lowest in energy is the ground state band and the other is the excited one. Much experimental and theoretical work has focused on this phenomenon and research is still 
ongoing. The review articles \unscite{coexistence}, \unscite{Wood} present much of this effort. The leading theoretical approach for shape coexistence up to date is via particle-hole excitations \unscite{HC}. This method is very successful in nuclei with $Z$ (proton number) being a magic number, but in open shell regions  the particle excitation mechanism cannot be the explanation (page 1486 of \unscite{coexistence}).

It is remarkable that experimental manifestations of shape coexistence (see Fig. (8) of \unscite{coexistence}) stop suddenly at a proton or neutron  harmonic oscillator magic number, \ie 2, 8, 20, 40, 70, 112, 168, 240,... The best examples are Fig. (9) from ref. \unscite{coexistence} and Fig. (3.10) from \unscite{Wood}. These figures present the energies of the various bands versus the mass or neutron number for the isotopic chains of $\ce{Hg}$ and $\ce{Sn}$. Two observations become obvious: firstly that the excited $0^+$ band of shape coexistence stops suddenly that 110 neutrons for $\ce{Hg}$ (just before the HO shell closure) and at 70 neutrons for $\ce{Sn}$ (HO shell closure) and secondly that the minimum of the excited band is at 104 for $\ce{Hg}$ and at 66 neutrons for $\ce{Sn}$, both of them being the mid shell regions of 82-126 and 50-82 shells respectively. Thus both magic numbers Proxy and HO should be considered for the explanation of shape coexistence. The irreps for the two sets for 28-50 shell are displayed in Table (1).

The previous hand-writing method is suitable for the calculation of the h.w. irrep, which corresponds to the ground state band. Another excited band can indeed be the outcome of particle-hole excitations. For instance, in $\ce{^{72}_{34}Se_{38}}$, proton excitation in the SU(3) glossary (not in the Shell Model) means that the 6 valence protons do not anymore occupy the orbitals $\ket{1},\ket{2},\ket{3}$, but possibly the $\ket{1},\ket{2},\ket{4}$. In such a case the proton irrep is (6,6). A pair of particles can even be excited into the next major shell, as it is assumed in the review articles \unscite{coexistence}, \unscite{Wood}. But an alternative irrep can also be extracted {\it without particle excitations}.This can be achieved by considering an alternative set of magic numbers. Instead of  using the 28-48 shell, one can use the 20-40 shell. Thus 8 additional particles participate in the $Q\cdot Q$ interaction. 

From a theoretical point of view one may wonder why, to use the harmonic oscillator magic numbers. The answer is that the algebraic $Q\cdot Q$ interaction is originally introduced through its
action in the space of orbitals with the same total number of quanta providing the HO magic numbers. Further insight can be gained by looking at the definition of the $q_m,m=-2,-1,0,1,2$ operators (chapter 4 of \unscite{Lipkin}). If $a_{j}, a_{j}^\dagger, j=z,x,y$ are the destruction and creation 3D HO operators (eq. 3.1 of \unscite{Harvey}) then:
\begin{eqnarray}\label{qq}
q_2=-{\sqrt{6}\over 2}(n_x-n_y)\mp {\sqrt{6}\over 2}i(a_x^\dagger a_y+a_y^\dagger a_x),\\
q_{\pm 1}=\mp\sqrt{3\over 2}(a_z^\dagger a_x+a_x^\dagger a_z\pm i(a_z^\dagger a_y+a_y^ \dagger a_z)),
q_0=2n_z-n_x-n_y.
\end{eqnarray}
The above are combinations of $a^\dagger a$ operators, so every time an $a$ operator destroys one quantum, the $a^\dagger$ places it in an other axis and finally the total number of quanta is conserved. This is the reason why they algebraic $q\cdot q$ interaction has matrix elements only among orbitals with the same $n$. But orbitals with the same $n$ occur among HO magic numbers (0, 2, 8, 20, 40, 70, 112, ...)

In the shell model it is the $l\cdot s$ interaction that has separated the 20-28 nucleons from the 28-40 ones, and as a result the nuclear shell 28-50 emerged. But there is a well known competition between the $l\cdot s$ interaction and the $q\cdot q$: in spherical nuclei $l\cdot s$ prevails, while in deformed nuclei $q\cdot q$ becomes dominant (section 7.3.2, exercise 7.6 of \unscite{Greiner}). Each type of interaction has its own magic numbers: the $l\cdot s$ creates the 28-50 shell, while the $q\cdot q$ applies in the 20-40 one. The final outcome in the intermediate region is the result of the competition among these two extremes.  A review article about magic numbers, in which the role of the harmonic oscillator magic numbers is pointed out, can be found in ref. \unscite{Sorlin}.
\section{ Harmonic oscillator irreps}

Using the shell 20-40, the valence protons and neutrons for  $\ce{^{72}_{34}Se_{38}}$ are 14 and 18 respectively. 
\begin{figure}
\begin{center}
\caption{GT triangle for 14 valence particles in U(10), for the h.w. irrep.}\label{GT3}
\bigskip
\includegraphics[scale=0.6]{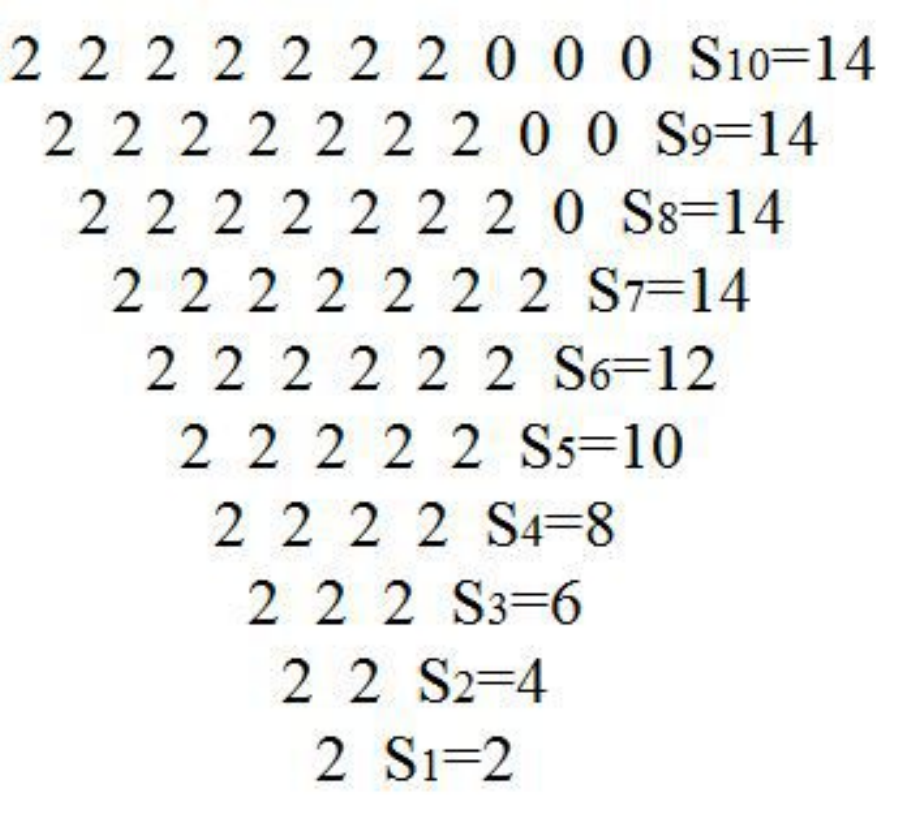}
\end{center}
\end{figure}
\begin{figure}
\begin{center}
\caption{GT triangle for 18 valence particles in U(10), for the h.w. irrep.}\label{GT4}
\bigskip
\includegraphics[scale=0.6]{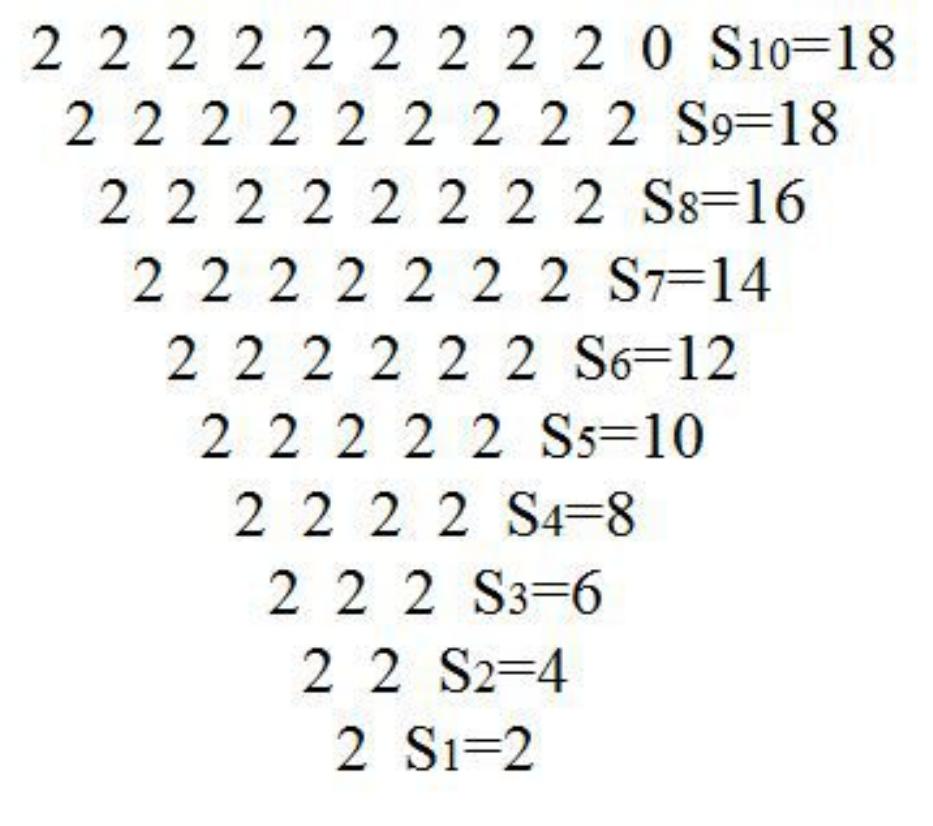}
\end{center}
\end{figure}
The GT triangle for the 14 valence protons is Fig. (\ref{GT3}). The weight vector is $\vec w=(2,2,2,2,2,2,2,0,0,0)$ and so the protons occupy the seven first orbitals $\ket{1},...\ket{7}$. Thus $\sum n_z=2\cdot 3+2\cdot 2+2\cdot 2+2\cdot 1+2\cdot 1+2\cdot 1+2\cdot 0=20$, $\sum n_x=2\cdot 0+2\cdot 1+2\cdot 0+2\cdot 2+2\cdot 1+2\cdot 0+2\cdot 3=14$ and $\sum n_y=2\cdot 0+2\cdot 0+2\cdot 1+2\cdot 0+2\cdot 1+2\cdot 2+2\cdot 0=8$. The U(3) quantum numbers are $[f_1,f_2,f_3]=[20,14,8]$ and $(\lambda_{2p}, \mu_{2p})=(6,6)$. Following the same procedure for the 18 valence neutrons the GT triangle is Fig. (\ref{GT4}) with $\vec w=(2,2,2,2,2,2,2,2,2,0)$ and $[f_1,f_2,f_3]=[20,20,14]$. So $(\lambda_{2N},\mu_{2N})=(0,6)$. The final highest weight irrep is
\begin{equation}
(\lambda_2,\mu_2)=(\lambda_{2p}, \mu_{2p})+(\lambda_{2N}, \mu_{2N}),
\end{equation}
which gives $(6,12)$ for this example. The corresponding $\gamma$ value for this irrep is $39.83^\circ$, so the nucleus is oblate.

\begin{table}\label{Table}
\caption{$(\lambda, \mu)$ of the highest weight irreps for two sets of magic numbers.}\label{irreps}
\begin{center}
\begin{tabular}{|c c c|}
\hline
Particle Number & HO & Proxy \\
\hline
28 & (10, 4) & (0, 0) \\
30 & (10,4) & (6,0) \\
32 & (12, 0)& (8, 2) \\
34 & (6, 6) & (12, 0)\\
36 & (2, 8) & (10, 4) \\
38 & (0, 6) & (10, 4)\\
40 & (0, 0) & (12, 0) \\
42 & (8, 0) & (6, 6) \\
44 & (12, 2)& (2, 8) \\
46 & (18, 0)& (0, 6) \\
48 & (18, 4)& (0, 0) \\
50 & (20, 4) & (0, 0) \\
\hline
\end{tabular}
\end{center}
\end{table}

\section{Conclusion}
The HO magic numbers are proposed as a complementary set with the proxy magic numbers, to derive the properties of the first excited $0^+$ band of shape coexistence. This complementary set, from a theoretical point of view, emerges naturally, as the set of orbitals among which the algebraic $q\cdot q$ interaction has non zero matrix elements. Therefore the h.w. irreps by both HO and proxy magic numbers should be combined to explain the intriguing ``shape coexistence". The proposed algebraic approach does not involve particle-hole excitations across the next major shell, thus it can also be used to predict open shell regions of shape coexistence.

\section*{Acknowledgments}

Helpful discussions with R. F. Casten and T. Mertzimekis are gratefully acknowledged. 
Work partly supported by the Bulgarian National Science Fund (BNSF) under Contract No. DFNI-E02/6.

\end{document}